\documentclass[11pt,fleqn,a4paper]{article} 
\usepackage[dvips]{graphicx,psfrag,overpic,color}
\usepackage{latexsym}
\usepackage{amssymb,amsmath,amsfonts,amsxtra,amstext}
\usepackage{rotating,tabularx,multicol,fancyhdr}
\usepackage[none]{hyphenat}
\usepackage[bookmarksnumbered, colorlinks, backref, bookmarks, breaklinks, linktocpage]{hyperref}
\hypersetup{
    bookmarks=true,        
    unicode=true,          
    pdftoolbar=true,       
    pdfmenubar=true,       
    pdffitwindow=true,     
    pdfsubject={Cisis10},  
    pdfnewwindow=true,     
    colorlinks=true,       
    linkcolor=red,         
    citecolor=red,         
    filecolor=blue,        
    urlcolor=blue,         
    a4paper=true
    }
\usepackage[all]{hypcap}
\begin{document}

\title{Trident, a new pseudo random number generator based on coupled chaotic maps\footnote{Submitted to the 3rd Int. Workshop on Computational Intelligence in Security for Information Systems (CISIS'10), Le\'{o}n (Spain), November 2010 }}
\author{A. B. Or\'{u}e, G. \'{A}lvarez, A. Guerra, G. Pastor,\\ M. Romera, and F. Montoya \\ \parskip=8pt
\small \emph{Applied Physics Institute, CSIC}\\
\small \emph{C/ Serrano 144, 28006-Madrid, Spain}
}

\date{}
\maketitle
\sloppy
\begin{abstract}
This article describes a new family of cryptographically secure pseudorandom number generators, based on coupled chaotic maps, that will serve as keystream in a stream cipher. The maps are a modification of a piecewise linear map, by dynamic changing of the coefficient values and perturbing its lesser significant bits.
\end{abstract}
\parskip=8pt
\sloppy

\section{Introduction} \label{sec:1}
A Pseudo Random Number Generator (PRNG) is an essential building block with many applications, including test data generation, Monte Carlo simulation techniques, spread spectrum communications, to name a few. But they are an essential part of any cryptosystem because of the security of many cryptographic systems depends on the generation of good pseudorandom sequences. The generated numbers will be used mainly as keystreams, initial vectors, private keys, and private signatures, destined to control or initialize cryptographic algorithms. However, the design of reliable pseudorandom generators remains an open problem in cryptology. Some \emph{de facto} standards that regarded as secure in the past have recently failed ~\cite{Klein08,Goldberg96,Gutterman06,Dorrendorf09}, other generators such as the BBS generator  ---which is one of the few PRNGs with proven security~\cite{BBS86}---\, are of little use for its slowness.

In 2000, the NESSIE (\emph{New European Schemes for Signatures, Integrity and Encryption}) project, was launched in Europe, as an open call, for the submission of cryptographic primitives. Unfortunately, all six stream ciphers submitted failed against cryptanalysis. In 2004, the eSTREAM project was launched as part of ECRYPT (\emph{European Network of Excellence in Cryptology}). eSTREAM issued a call for stream cipher primitives. As a result, in 2008 seven finalists where pre selected (four in software and 3 in hardware), but currently it has not yet been possible to decide which of them deserves to be a standard.

The link between cryptography and chaotic systems continues to be an active research field. Many researchers agree that the interaction of these areas can be mutually beneficial. Many tools for analyzing chaotic systems have also served as tools in cryptanalysis of many systems and for studying and improving the design of others~\cite{Alvarez00,Alvarez06,Orue08}.

This paper presents a family of pseudo random number generators, that consist of several coupled chaotic maps, mutually perturbed, that will serve as keystream in a stream cipher.

\section{Family of Pseudorandom Generators Based on the Combination of Chaotic Maps}

The proposed  family of pseudorandom generators  is based on the combination of the sequences generated by several coupled basic pseudorandom generators, through a one-way function. Each one consists of a chaotic map, cryptographically secure itself, which has a high entropy.

Every map has a limited number of states, and therefore its period of repetition is also  limited; in agreement to the word length of the language with which it is programmed, that in turn depends on the word length of the hardware that is in use.

The combination of several sequences by a one-way function has two aims. The first one is to increase the number of states of the system, with the consequent increase of the period of the repetition, the increase of the entropy, and the increase of the number of keys. The second objective is to increase the security. Indeed, when mixing multiple streams so that the size of the output word is less than the sum of the sizes of the input words, it is extremely difficult to make an individualized analysis of the sequences generated by each chaotic map and, hence, to mount a cryptanalytic attack.

The method of combination chosen consists of the bitwise XOR of the numbers generated by several chaotic maps. The number of maps must be chosen depending on the application and on the word length of the software with which it is programmed.

It was used a mixing of arithmetical operations and operations oriented to bit, because this serves to avoid the purely algebraic attacks and the purely bit oriented attacks. The mixing of a variety of operations, as  algebraic and bit manipulations, prevents the mathematical behavior of the scheme from being shaped.

In addition, only very efficient operations are used: arithmetical operations module the word size of the compiler, bitwise boolean operations, and displacements of bits, all of them of easy implementation in hardware or software. All these operations combined in the proposed generator contribute to a great mathematical complexity together with a high computational efficiency.

\subsection{Chaotic Map With Dynamical Variation of Coefficients and Perturbation of the Least Significant Bits}
Assume there are one-dimensional chaotic maps $F(x_t, a, c, s)$ which consists of the modification of a piecewise linear chaotic map by means of the dynamic increase of coefficients and the perturbation of the least significant bits of the samples, where $\{x_t\}$  denote the chaotic orbit and $a_t, c_t$ are the control parameters of the system, which is defined by:
\begin{align}
x_t&= (a_t\, x_{t-1} +c_t)\!\!\!\! \mod m \oplus ((a_t\, x_{t-1} +c_t)\!\!\!\! \mod m)\! \gg s, \label{equ:sencillo} \\
a_t&= (a_{t-1}+ \Delta a) \mod m, \label{equ:increment_a}\\
c_t&= (c_{t-1}+ \Delta c) \mod m, \label{equ:increment_c}
\end{align}
where $t$ denotes the time; $n$ is the number of bits of precision used; $m = 2^n$ is the maximum number of values that $x_t$ can take ; $a_t, c_t$ are the coefficients whose values suffer respectively an increase of $\Delta a, \Delta c$ with every iteration of the system;  $x_0, a_0, c_0$ are the initial conditions; $s$ is a constant $0\le s \le n$; $\gg\! s$ is the right-shift operator in the C/C++ language, that is equivalent to a division by $2^s$ followed by the \texttt{floor} operation, i.e. $y\gg s =\lfloor y/2^s \rfloor $.

The innovation of this chaotic map is that it comprises three interlinked operations of different nature, the first one is a linear operation $x_t= (a_t\, x_{t-1} +c_t)\!\!\! \mod m$, peculiar to the algebraic pseudorandom generators, while the second one is the bitwise XOR, and the third one is the right-shift  $\gg s$, both peculiar to the shift registers pseudorandom generators. The coefficients $a_t$ and $c_t$ are changed dynamically over time.

This map may be considered a modification of a sawtooth piecewise linear map incorporating two novelties, the first one is the \emph{dynamical increment of the coefficients}, which hides any observable regularity; and the second one is the \emph{perturbation of less significant bits} attained by the bitwise XOR addition of the right-shifted bits, with the aim of increasing the entropy of the generated number sequence.

Drawing a two-dimensional graph of the points of the chaotic orbit $\{x_{t}\}$ as a function of $\{x_{t-1}\}$, yields the so-called return map function, which in simple cases can reconstruct the value of the parameters of a chaotic orbit.

\subsubsection{Plain Sawtooth Piecewise Linear Map} \label{ej:1}
The simplest form of the generator takes place when $\Delta a = \Delta c = 0$ and $s=n$. When $s=n$, Eq.~\eqref{equ:sencillo}  is reduced to $x_t= (a_t\, x_{t-1} +c_t)\!\!\! \mod m$, because a  $n$ bits right-shift of a number coded with $n$ bits will be always equal to zero. In such a case, the resulting map is a sawtooth piecewise linear map which coincides with the lineal congruential generator, whose lack of security is well known~\cite{Knuth97,Bellare97}.

The maximum value of the period $p=m$ of the generated sequence, is reached  when $a_0\!\!\mod 4=1$ and $c_0\!\!\mod 2 = 1$~\cite{Knuth97}.

\begin{figure}[t]
\capstart
\begin{center}
\begin{overpic}[width=58mm]{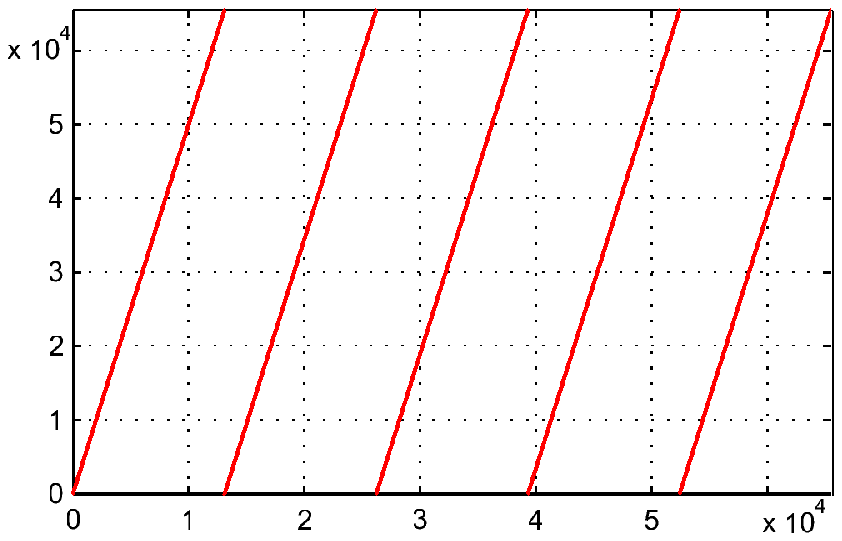}
\put(0, 35){\small $x_{t}$ }
\put(53, 0){\small $x_{t-1}$ }
\end{overpic}
\caption{Return map of the mapping $x_t=(5x_{t-1}+1)\!\!\mod m$,
with $m=2^{16}$.}
\label{fig:retorn0}
\end{center}
\end{figure}

In this case the return map of the function of Eq.~\eqref{equ:sencillo} is reduced to a sawtooth, with so many steps as the value of $a_0$ and with an ordinate at the origin equal to $c_0$. Figure~\ref{fig:retorn0} illustrates an example for $a_0=5$, $c_0=1$, and $m=2^{16}$.

The map for these parameter values is bijective, hence invertible, and subject to algebraic cryptanalysis. It lacks security because the parameter value may be deduced by inspection of the return map.

The entropy of the generated number sequence is very low, the lesser significant bits of the numbers are not random; for instance, the sequence of the least significant bit of the numbers, generated by the map of the example of Fig.~\ref{fig:retorn0}, is just $\{1,0,1,0,1,0,1,0, \ldots\}$, the sequence of the second less significant bit is  $\{1,1,0,0,1,1,0,0, \ldots\}$, and so on. Only the higher significant bits can be considered random. This generator fails to pass most of randomness test suites.

\subsubsection{Sawtooth Piecewise Linear Map with Dynamical Variation of Coefficients} \label{ej:2}
The firs step toward a satisfactory level of security consists of the modification of the map described in Sect.~\ref{ej:1} by incrementing the value of parameters $a_t$ and $c_t$ at each iteration of the system, according to Eqs.~\eqref{equ:increment_a} and~\eqref{equ:increment_c}. The maximum value of the period $p=m$, of the generated sequence, is reached  when $s=n$, $a_0\!\!\!\mod 4=1$, $\Delta a\!\!\! \mod4=0$, $c_0\!\!\!\mod 4 = 1$, and $\Delta c\!\!\! \mod4=0$.

\begin{figure}[h]
\capstart
\begin{center}
\begin{overpic}[width=58mm]{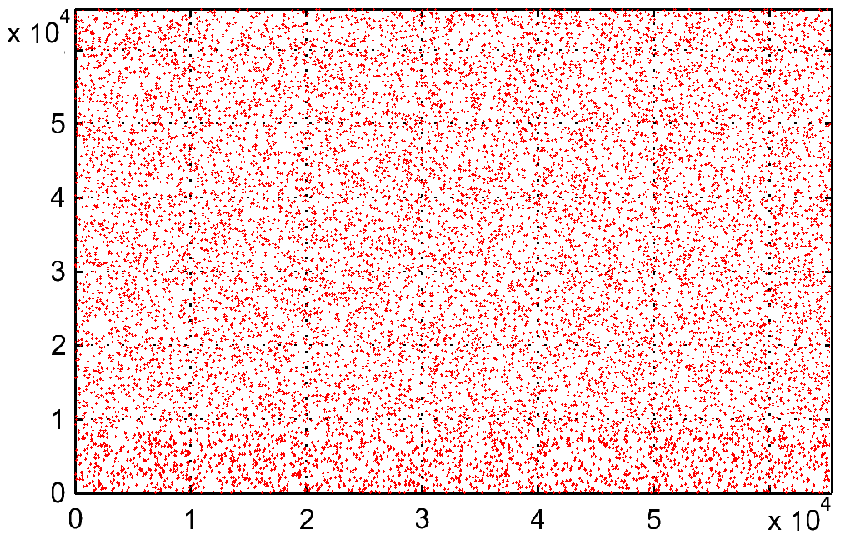}
\put(1, 35){\small $x_{t}$ }
\put(53, 0){\small $x_{t-1}$ }
\put(50,65){(a)}
\end{overpic}
\begin{overpic}[width=58mm]{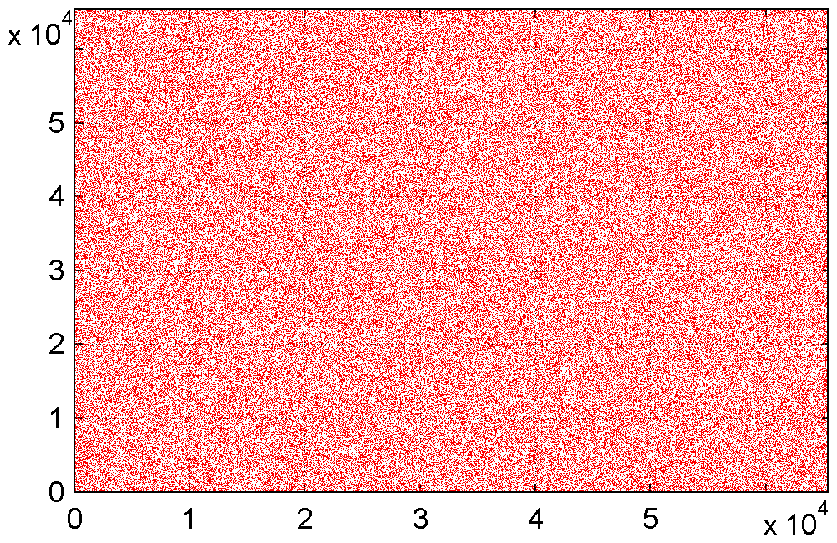}
\put(0, 35.5){\small $x_{t}$ }
\put(54, 0){\small $x_{t-1}$ }
\put(50,65){(b)}
\end{overpic}
\caption{Return map of the mapping of Eq.~\eqref{equ:sencillo}, for $a_0=5$, $\Delta a=4$, $c_0=1$, $\Delta c=4$, $n=16$ and $m=2^{16}$;  (a) $s=n=16$; (b) $s=8$.}
\label{fig:retornos}
\end{center}
\end{figure}

The corresponding return map is depicted in Fig.~\ref{fig:retornos}\,(a), for $a_0=5$,\; $\Delta a =4$,\; $c_0=1$,\; $\Delta c =4$,\; $s=n=16$\; and $m=2^{16}$. It can be observed that the return map does not supply any information about the parameters nor their increments; the number of points in it is the same as in Fig.~\ref{fig:retorn0} but they are scattered across the whole return map area. But the mapping is still bijective and invertible, hence subject to algebraic cryptanalysis and the improvement of security is limited.

The dynamical variation of coefficients, incrementing their value at each iteration, acts as a counter, changing permanently the estate of the map, thus preventing the existence of fixed points or short periods of the generated sequence. The maximum repetition period of the coefficients is $2^{n-2}$ which is reached when $\Delta a$ or $\Delta c$ have the least possible common prime factors with $m$, what is achieved for $\Delta a\!\!\! \mod8 = \Delta c\!\!\! \mod8=4$ when $m=2^n$.

\subsubsection{Full Version of the Chaotic Map with Dynamical Variation of Coefficients and Perturbation of the Least Significant Bits}  \label{ssec:todo}
With the full version of the mapping defined by Eqs.~\eqref{equ:sencillo},~\eqref{equ:increment_a} and \eqref{equ:increment_c}, the optimum entropy and security is reached. The range of observed periods varies between $p=2^{1.5n}$ and  $p=2^{2n}$ for different parameters with values $s=n$, $a_0\!\!\!\mod 4=1$, $\Delta a\!\!\! \mod8=4$, $c_0\!\!\!\mod 4 = 1$, and $\Delta c\!\!\! \mod8=4$.

Figure.~\ref{fig:retornos}\,(b) illustrates the return map for the following parameter values: $a_0=5$, $c_0=1$, $\Delta a = 4$, $\Delta c = 4$, $s=8$, $n=16$ and $m=2^{16}$.

Note that this map contains much more points than the one corresponding to the Fig.~\ref{fig:retornos}\,(a), because the period of the sequence is much longer (for clarity only the first $N=4m=2^{18}$ points are represented).

The full version of the mapping defined by Eqs.~\eqref{equ:sencillo}, \eqref{equ:increment_a} and \eqref{equ:increment_c}, is not injective because each element of the domain is mapped to several elements of its codomain, and each element of the codomain may have several preimages. Hence it is impossible to invert the mapping.

The number $M$ of different samples in the generated sequence obeys to the birth day problem formula:
\begin{align}\label{equ:cumpleannos}
M = m - m \left(\frac{m-1}{m}\right)^N
\end{align}
being $N$ the total number of generated samples. The statistical behavior is identical to one of a pure random source.

The sequences generated by the full version of the mapping (with a word size of $n=64$ bits) pass successfully the randomness test suites of the American National Institute of Standards and Technology,  NIST SP~800-22~\cite{SP800-22}, as well as the more stringent \emph{Diehard} from Marsaglia (1995) and even the new and yet more stringent Tuftests of Marsaglia and Tsang of 2002~\cite{Marsaglia02}.

It was found that the best values of $s$ to pass the mentioned tests,   ranged from 20 to 44 right-shift places, when a word size of $n=64$ bits was used.

A fundamental achievement of this scheme is the correction of the lack of entropy of the less significant bits of the sequence generated by a sawtooth piecewise linear map, that is corrected by the bitwise XOR with the most significant bits.


\subsection{Trident Combined Generator}
The full version of the mapping defined by Eqs.~\eqref{equ:sencillo}, \eqref{equ:increment_a} and \eqref{equ:increment_c}, may constitute itself an excellent random number generator; but there are two remaining problems to be considered.

The first one is that if the word size used is limited, the maximum period will be limited and may result too short for certain applications, such as stream cipher of video signals.

The second problem is that the output generated numbers are the same that are used to calculate the next sample. An opponent may try to mount an algebraic attack, that could be feasible if the attacker have access to a large computer facility.

\begin{figure} [t]
\capstart
\begin{center}
\includegraphics[width=125mm]{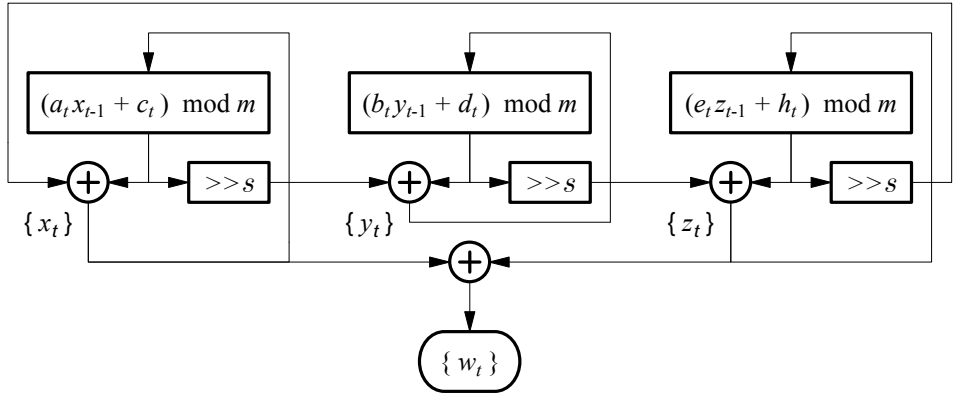}\\
\caption{Trident combined generator.} \label{fig:trident}
\end{center}
\end{figure}

To avoid these troubles, a joint combination of three individual chaotic maps, with dynamical variation of coefficients and perturbation of the least significant bits, named \emph{Trident}, is proposed. One of the maps to be combined will remain completely hidden, while the sequences generated by the other two are combined by the bitwise XOR addition, to form the output of the joint generator. This architecture is depicted in Fig.~\ref{fig:trident}.

Note that the way of combining the three individual generators is the chained perturbation of the less significant bits. Instead of perturbing the less significant bits of each map through the bitwise XOR of the own right-shifted bits, they are perturbed by the right-shifted bits of the nearby map in a cyclic chained manner.

In this way, two goals are attained: first, the period of the joint generator is increased to the least common multiple of the periods of the three individual generators; second, the analysis of the output sequence is useless for determining the system parameters.

The equations defining the Trident generator are:
\begin{align}
w_t&= x_t \oplus z_t,\\
x_t&= (a_t\, x_{t-1} +c_t)\!\!\!\! \mod m \oplus ((e_t\, z_{t-1} +h_t)\!\!\!\! \mod m)\! \gg s,  \\
y_t&= (b_t\, y_{t-1} +d_t)\!\!\!\! \mod m \oplus ((a_t\, x_{t-1} +c_t)\!\!\!\! \mod m)\! \gg s, \\
z_t&= (e_t\, z_{t-1} +h_t)\!\!\!\! \mod m \oplus ((b_t\, y_{t-1} +d_t)\!\!\!\! \mod m)\! \gg s,\\
a_t&= (a_{t-1}+ \Delta a) \mod m; \,\quad c_t= (c_{t-1}+ \Delta c)\mod m,\\
b_t&= (b_{t-1}+ \Delta b) \mod m; \,\quad d_t= (d_{t-1}+ \Delta d) \mod m,\\
e_t&= (e_{t-1}+ \Delta e) \mod m; \,\quad h_t= (h_{t-1}+ \Delta h) \mod m,
\end{align}
were $w_t$ is the output sample of the Trident generator at  the moment $t$; $x_t, y_t, z_t$ are the output samples of the three generators; $a_t, c_t, b_t, d_t, e_t, h_t$ are the correspondent coefficients; and $\Delta a, \Delta c, \Delta b, \Delta d, \Delta e, \Delta h$ are the increments of the respective coefficients. All the coefficients and the respective increments should be chosen of different values, to warrant that the sequences generated by each generator are unique.

The Trident generator key is formed by the whole coefficients of the individual maps. Due to the limitations imposed to the coefficient values (see Sect.~\ref{ssec:todo}) the number of bits with which the coefficients are coded is $n-2$ and the number of bits with which the increments of the coefficients are coded is $n-3$. If the word size is
64 bits, the total amount of key bits is $3\times 62 + 3\times 61= 369$,
considering that all coefficients must be different, one must accept that the effective number of bits is $368$. If one considers the initial values $x_0$, $y_0$ and $z_0$ as part of the key, we must add $3 \times 64=192$ more bits to the key length, hence the total amount of different keys rise to $2^{560}$.  As the right-shift value allows for a very small variation, it is not worth its consideration as part of the key. It is better to fix its value to $32$ bits.

The evident form to attack the system is the brute force but the huge number of different keys prevents such attack. The algebraic attack is reasonably unlikely due to the impossibility of learning the internal state of the generator.

Large number of sequences were generated by the Trident combined generator with a word size of 64 bits, programmed in C99. All of them passed with success the randomness test suites of the NIST SP~800-22, as well as the \emph{Diehard} from Marsaglia (1995) and  the Tuftests of Marsaglia and Tsang of 2002.

The performance of Trident in an Intel \emph{Core2 Duo} with OS Windows32 is about one clock cycle/bit. This speed is in the range of the finalists of the eSTREAM project.

Different versions may be designed using more than three coupled chaotic maps and perturbed in the same way and different word sizes.  For instance, to compensate the smaller periods attainable with an  architecture of only 32 bits, a five chaotic maps can be used, in this way two completely different sequences, with the same repetition period, could be generated.

\section{Conclusion}
 A fast, cryptographically secure pseudorandom number generator, has been described, based on the combination of three coupled chaotic maps. The maps are sawtooth piecewise linear maps with dynamical variation of coefficients and perturbation of the least significant bits. Its output is unpredictable. The version with a word size of 64 bits has a repetition period length in excess of $2^{198}$ bits. The generated sequence passes successfully the most stringent randomness test suites. The attained performance is about one clock cycle per generated bit.

\subsubsection*{Acknowledgement}
This work was supported by  Ministerio de Ciencia e Innovaci\'{o}n of Spain projects CUCO MTM(2008-02194) and TEC2009-13964-C04-02; and by CDTI (Ministerio de Industria, Turismo y Comercio) in collaboration with Telef\'{o}nica I+D, project \-SEGUR@ (CENIT 2007-2010).

\end{document}